\pdfoutput=1

\documentclass[10pt,journal,a4paper,final,oneside,twocolumn]{IEEEtran}
\usepackage{times}
\usepackage{graphicx}
\usepackage{cite}
\usepackage{color}
\usepackage{psfrag}
\usepackage{subfigure}
\usepackage{amssymb}
\usepackage{epsfig}
\usepackage{pifont}
\usepackage{amsmath}
\usepackage{array}
\usepackage[T1]{fontenc}
\usepackage{comment}
\usepackage{acronym}
\usepackage{multirow}
\usepackage{epsfig}
\usepackage{booktabs}

\begin{document}

\title{Deep Joint Source-Channel Coding for DNA Image Storage: A Novel Approach with Enhanced Error Resilience and Biological Constraint Optimization}


\author{\IEEEauthorblockN{
                Wenfeng Wu, \emph{Student~Member, IEEE},
			Luping Xiang, \emph{Member, IEEE},
   			Qiang~Liu, \emph{Member, IEEE},
			and~Kun~Yang}, \emph{Fellow, IEEE}
			\vspace{-0.5 cm}\\

\thanks{This work was supported in part by the Natural Science Foundation of China under Grant 62301122 and Grant 62071101; in part by the Fundamental Research Funds for the Central Universities under Grant ZYGX2019J001; in part by the Sichuan Science and Technology Program under Grant 2023NSFSC1375. \textit{(Corresponding author: Luping Xiang.)}}
        \thanks{Wenfeng Wu and Luping Xiang are with the School of Information and Communication Engineering, University of Electronic Science and
        Technology of China, Chengdu 611731, China, email: wenfengwu@std.uestc.edu.cn, luping.xiang@uestc.edu.cn. }
        \thanks{Qiang Liu is with the Yangtze Delta Region Institute (Quzhou), University of Electronic Science and Technology of China, Quzhou, Zhejiang 324000, China, email: liuqiang@uestc.edu.cn.
        }

        \thanks{Kun Yang is with the School of Computer Science and Electronic Engineering, University of Essex, CO4 3SQ Colchester, U.K, and also with the School of Information and Communication Engineering, University of Electronic Science and Technology of China, Chengdu 611731, China, e-mail: kunyang@essex.ac.uk).}
	}

\maketitle

\begin{abstract}
In the current era, DeoxyriboNucleic Acid (DNA) based data storage emerges as an intriguing approach, garnering substantial academic interest and investigation. This paper introduces a novel deep joint source-channel coding (DJSCC) scheme for DNA image storage, designated as DJSCC-DNA. This paradigm distinguishes itself from conventional DNA storage techniques through three key modifications: 1) it employs advanced deep learning methodologies, employing convolutional neural networks for DNA encoding and decoding processes; 2) it seamlessly integrates DNA polymerase chain reaction (PCR) amplification into the network architecture, thereby augmenting data recovery precision; and 3) it restructures the loss function by targeting biological constraints for optimization. The performance of the proposed model is demonstrated via numerical results from specific channel testing, suggesting that it surpasses conventional deep learning methodologies in terms of peak signal-to-noise ratio (PSNR) and structural similarity index (SSIM). Additionally, the model effectively ensures positive constraints on both homopolymer run-length and GC content.
\end{abstract}

\begin{IEEEkeywords}
DNA Storage, Deep Learning, Joint Source-Channel Coding, Biological Constraints.
\end{IEEEkeywords}

\section{Introduction}

\begin{figure}[htbp]
  \centering
    \includegraphics[width=0.5\textwidth]{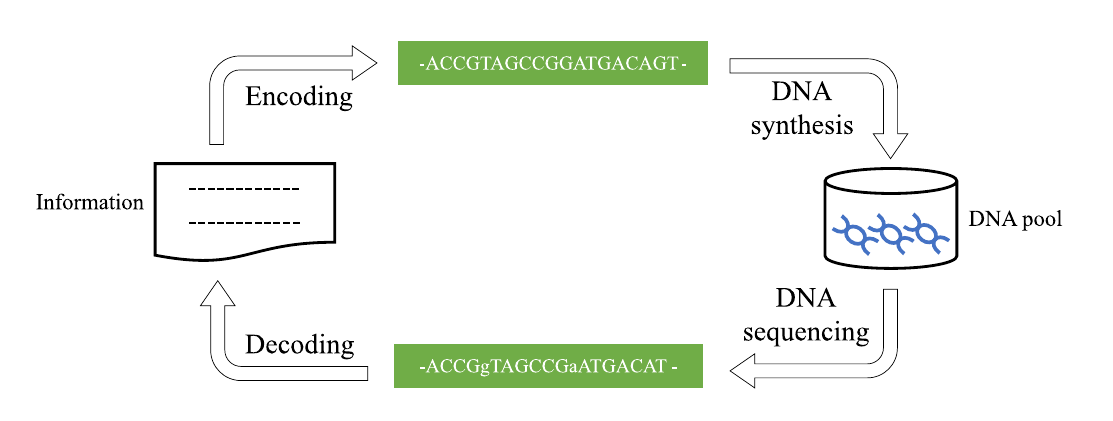}
    \caption{The process of storing digital information on DNA.}
\label{fig.process}
\end{figure}

Over recent years, DNA storage has emerged as an attractive approach for data storage due to its key benefits, such as its high capacity, longevity, and substantial information density\cite{2001Long}. The enduring significance of DNA as the foundation of life underscores its potential as a robust storage medium. As illustrated in Fig.~\ref{fig.process}, this storage mechanism entails the transformation of digital data into DNA nucleotide sequences, the synthesis of the corresponding DNA molecules, and their subsequent storage within a DNA pool. The retrieval of the stored information necessitates the sequencing of these DNA molecules and the subsequent decoding to restore the original digital form\cite{2016A}. Generalized DNA storage can accommodate a diverse range of data types including text, images, and audio. Such information is first translated into binary or other numeral system representations and then encoded with error correction capabilities prior to DNA storage\cite{Ceze2019MolecularDD,2020research}. The feasibility of DNA as a storage medium has been a topic of discussion among computer scientists and engineers since the 1960s\cite{neiman1964some, baum1995building}, and significant breakthroughs have occurred with the recent progress in DNA synthesis and sequencing technology\cite{2012Next,2013Towards,2015Robust,2017DNA,2017CRISPR,2018Random,Koch2019ADS}. For instance, in 2012, Church et al. \cite{2012Next} managed to store $659$ $\mathrm{KB}$ of digital data in DNA, which was later improved by Goldman et al, who stored $739$ $\mathrm{KB}$ of data in 2013 \cite{2013Towards}. As research continued, DNA storage systems became more advanced, and in 2019, Erlich et al.\cite{Koch2019ADS} accomplished the stable replication and inheritance of a DNA blueprint via 3D printing of a Stanford bunny encoded with genetic information.

DNA coding, a pivotal aspect of DNA storage, targets the high-density and high-fault-tolerant storage of information. This entails three main phases: data compression for storage efficiency enhancement, error correction coding for increased data recovery accuracy, and nucleotide coding for the transformation of digital information into base sequences \cite{Ceze2019MolecularDD,2020research}. Presently, a majority of the compression and error-correction algorithms find their roots in classical coding theory. Huffman and fountain codes are frequently employed for data compression tasks\cite{2013Towards,2019Author,2017DNA,anavy2019data}, while RS codes, LDPC, and BCH codes have established themselves as predominant algorithms for error correction\cite{2021An,2015Robust,2018Random,2016Forward,anavy2019data}. In the realm of DNA storage, there are three fundamental nucleotide coding methodologies: binary coding\cite{zhirnov2016nucleic}, ternary coding\cite{2013Towards}, and quaternion coding\cite{2019Author}.

In addition to the conventional algorithms rooted in coding theory, recent research endeavors have started to integrate deep learning techniques to enhance storage efficacy and mitigate DNA-related expenses \cite{2021generative,Zhang2021HighdensityIS,2021Deep}. With the proven efficacy of deep learning in image compression\cite{2015Compression,2016D3,balle2016end}, there lies an opportunity to harness neural networks in the domain of DNA storage, aspiring for superior storage densities. Notably, autoencoders—a specific neural network architecture designed for dimensionality reduction and subsequent reconstruction—have the potential to represent original data in a compact form through unsupervised learning techniques.

Historically, DNA storage architectures have adopted compartmentalized approaches for both source and channel coding. In such systems, the purpose of source compression is to eliminate redundant information, ensuring the accurate reconstruction of the original data during decoding. Conversely, channel coding introduces supplementary check bits to the foundational sequence, bolstering the transmission resilience of sequences in the presence of channel noise. This approach, however, contrasts with source compression, and their bifurcated nature can result in less-than-ideal outcomes due to the inherent challenges of striking a balance\cite{Zhai2005JointSC}. Contemporary research suggests that joint source-channel coding (JSCC) strategies, especially those underpinned by deep learning, outperform the traditional isolated coding methodologies\cite{2019deep,2021bandwidth}.  For example, in 2019, Bourtsoulat et al.\cite{2019deep} introduced a scheme based on deep JSCC for image transmission over wireless channels, demonstrating its performance advantage over separate schemes.

To the best of our knowledge, there exists no literature that presents a JSCC approach tailored for DNA storage. Addressing this research void, our paper endeavors to integrate deep learning, specifically aiming at the formulation of a JSCC framework for DNA-based image storage. The predominant structure employs a convolutional neural network (CNN), renowned for its proficiency in feature extraction from images\cite{Jogin2018FeatureEU}. This methodology possesses the potential for broader adaptations, potentially serving as a universal DNA storage technique. Recognizing the pivotal role of image data in storage paradigms, this work provides an exemplary instance of neural network utilization in the context of DNA storage. The salient contributions of our investigation are delineated as follows:

\begin{itemize}
\item In our work, we seamlessly integrated deep learning into the DNA storage pipeline, supplanting traditional coding mechanisms with neural networks to conceive a JSCC system anchored on these networks. The method we propose, termed DJSCC-DNA, distinguishes itself from standard practices by utilizing a neural network in the coding phase, facilitating a direct translation of image pixel intensities to nucleotide sequences. This cultivates a comprehensive framework for image conservation in DNA.

\item Furthermore, we have incorporated DNA PCR amplification into our network architecture to enhance the fault tolerance of the DNA chains, which can be viewed as a form of information retransmission. Before decoding, we preprocess this retransmitted information. Considering the biological constraints, we have redesigned the loss function to incorporate both image restoration quality and biological requirements as targets for model optimization.

\item Our simulation results show that the proposed scheme effectively caters to the dual requirements of high-quality image resolution restoration and adherence to biological constraints. Specifically, at a compression ratio of $R=1/4$, our proposed scheme outperforms the conventional neural network-based scheme, VAEU-QC, by $5.10$ $\mathrm{dB}$ in PSNR and $0.439$ in SSIM. Additionally, our proposed loss function proves to be effective in controlling homopolymer run-length and GC content within the nucleotide sequence.
\end{itemize}

The structure of the remainder of this paper is as follows: Section II outlines the DJSCC-DNA system model; Section III provides details on the DNA storage based end-to-end design; numerical results are presented in Section IV; and the paper concludes in Section V.

\section{System Model}
\label{sec.System Model}

\begin{figure*}[!htbp]
  \centering
    \includegraphics[width=1\textwidth]{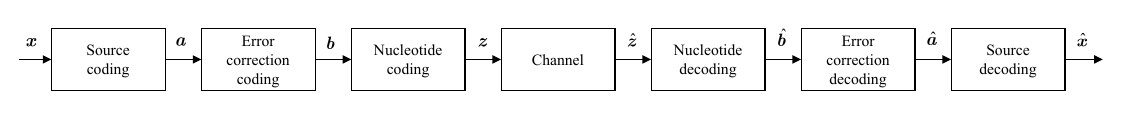}
    \caption{The conventional system model.}
\label{fig.traditional}
\end{figure*}

The conventional DNA storage system, illustrated in Fig.~\ref{fig.traditional}, is comprised of an encoding process (encompassing source coding, error correction coding, and nucleotide coding) and a decoding process (involving nucleotide decoding, error correction decoding, and source decoding). The encoded DNA strand, denoted as $\pmb{z}$, is synthesized and stored in a DNA pool. To recover the original data $\pmb{x}$, the target DNA strand, designated as $\pmb{\hat{z}}$, is extracted and sequenced, and subsequently decoded to reconstruct the data. This progression from $\pmb{z}$ to $\pmb{\hat{z}}$ can be conceptualized as a channel. Notably, PCR technology is employed to amplify $\pmb{z}$ during the DNA strand extraction process. However, this step is frequently overlooked in traditional coding approaches, and the specifics are typically not factored into the coding.

\begin{figure*}[!htbp]
  \centering
    \includegraphics[width=1\textwidth]{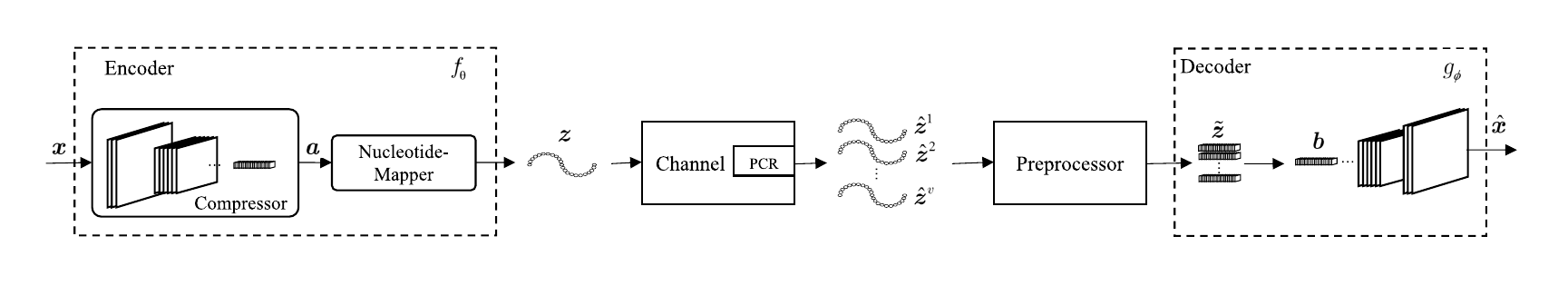}
    \caption{The DJSCC-DNA system model.}
\label{fig.systemmodel}
\end{figure*}

Our deep learning-based proposal presents a JSCC scheme for DNA image storage, as depicted in Fig.~\ref{fig.systemmodel}. This proposed scheme succinctly integrates an encoder, channel, preprocessor, and decoder, collectively creating a comprehensive end-to-end DNA storage framework. In contrast to conventional encoding procedures, DJSCC-DNA utilizes a CNN for the encoder, achieving source coding, error correction coding, and nucleotide coding functionalities. The decoder follows a similar principle. The PCR amplification process is integrated into the channel, regarded as a form of information retransmission aimed at enhancing data recovery precision. The preprocessor transforms the channel output into a specific format type, $\pmb{\tilde{z}}$, which can be computed and trained by the neural network. The channel and preprocessor, devoid of parameters, are considered non-training layers. The entire DJSCC-DNA model is trained in an integrated manner, allowing for efficient encoding and decoding of images once the training has been completed.

\subsection{Encoder}
\label{sec.Encoder}
An input image, dimensionally defined as $H$ (height) × $W$ (width) × $C$ (channels), can be mathematically represented by a matrix $\pmb{x}\in\mathbb{R}^{H\times W\times C}$. Within this context, the total pixel count, symbolized as $n$, is quantified by the relation $n=H\times W\times C$. The encoder transposes these input values into a nucleotide sequence of the channel input symbols $\pmb{z}\in \mathbb{Z}^k$, using the encoding function $f_\theta:\mathbb{R}^{H\times W\times C} \xrightarrow{} \mathbb{Z}^{k}$. Here, $k$ signifies the nucleotide sequence size and $\theta$ denotes the parameter set of the encoder. This encoding process can be expressed as:
\begin{align}\label{eq.enc}
\pmb{z}=f_{\theta}(\pmb{x})\in\mathbb{Z}^k.
\end{align}

The encoder includes two critical components: the compressor and the nucleotide-mapper. In the encoding phase, compressing the image will inevitably lead to information loss in terms of image details, and mapping pixel values to nucleotides introduces quantization errors that further contribute to information loss.

$\emph{1)}$ Compressor: This element is responsible for creating a vector representative of the potential portrayal of the input image. 
The image processing pipeline initially applies a normalization layer to adjust the pixel values of the input image within the range $[0,1]$. Following this, a CNN is engaged to derive pertinent features from the image. This network incorporates convolutional layers, Batch Normalization layers, and PReLU activation function layers. Structures of this sort are frequently employed to extract structural image features for compression, as highlighted in prior studies \cite{2019deep, 2018an}. The compressed vector $\pmb{a} \in \mathbb{R}^k$ produced from the image needs to be converted into a nucleotide sequence, as it cannot be stored directly.

\begin{figure}[htbp]
  \centering
    \includegraphics[width=0.25\textwidth]{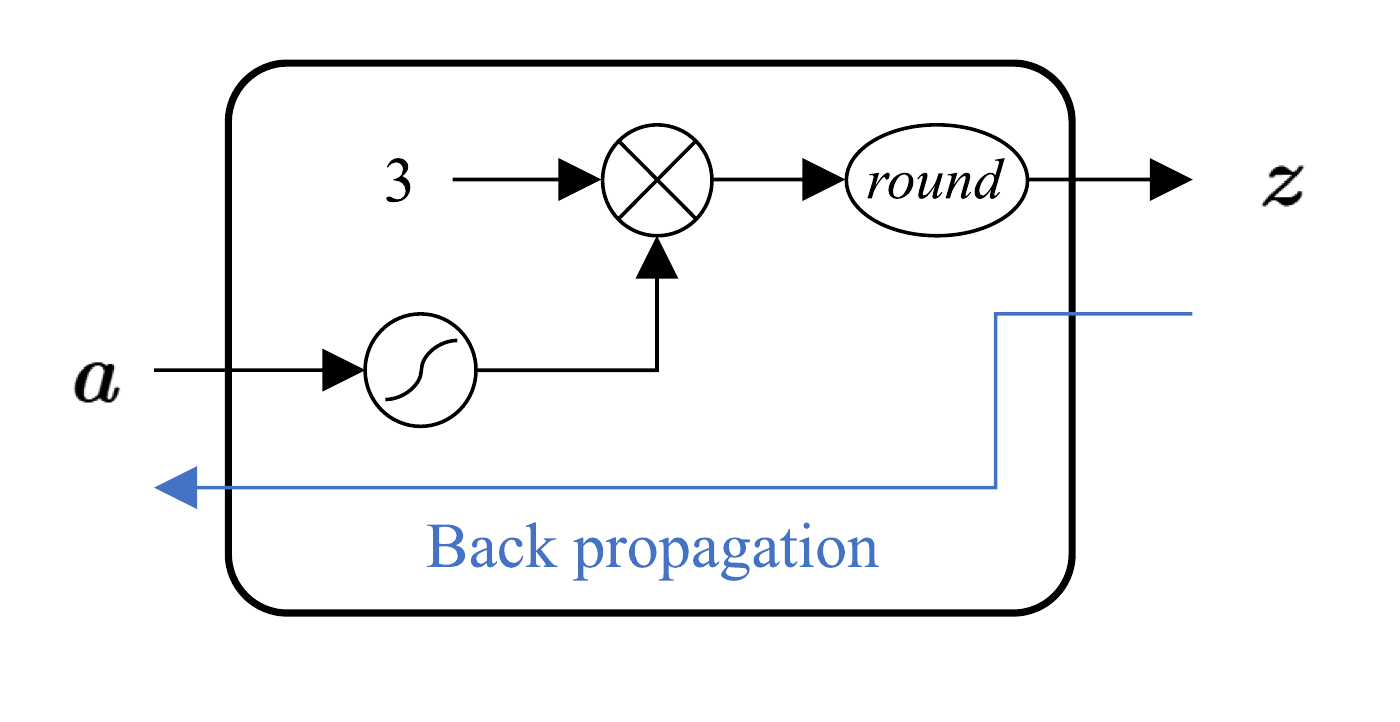}
    \caption{The structure of nucleotide-mapper.}
\label{fig.mapper}
\end{figure}

$\emph{2)}$ Nucleotide-mapper: This function serves to convert the output values, $\pmb{a}$, from the compression stage into a nucleotide sequence, as depicted in Fig.~\ref{fig.mapper}. Each resulting unit from $\alpha$ undergoes scaling within the $(0,1)$ range via the activation function $Sigmoid(\cdot)$. After amplification by a factor of three, these units are rounded, yielding values within the set $\{0,1,2,3\}$. These values are then mapped to the nucleotide representations $\{A,C,G,T\}$. An important consideration is the inherent non-differentiability of the round function in the nucleotide mapper. This characteristic impedes gradient backpropagation during deep neural network training. To address this limitation, we adjust the gradient value to 1.

\subsection{Channel model and preprocessor model}
\label{sec.Channel Model}

\begin{figure}[htbp]
  \centering
    \includegraphics[width=0.5\textwidth]{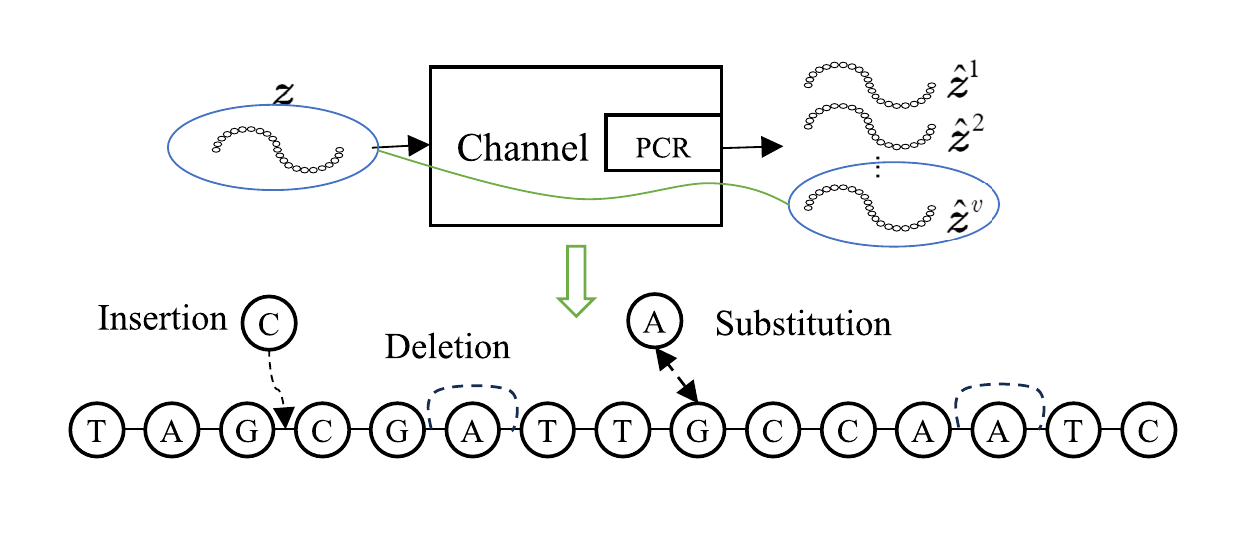}
    \caption{The structure of the channel model.}
\label{fig.propor1}
\end{figure}

The storage and retrieval of the encoded nucleotide sequence $\pmb{z}$ comprise the subsequent stages:

$\emph{1)}$ DNA Synthesis: This procedure pertains to the method of constructing a DNA sequence by arranging nucleotide units in a particular sequence. Data information is typically contained in short DNA strands, which exhibit superior error performance compared to longer DNA strands. For data retrieval, primers are attached to the front of each short DNA strand before they are fused into a longer DNA strand. These DNA strands are then stably preserved within a DNA pool.

$\emph{2)}$ DNA PCR Amplification: During DNA synthesis, each short DNA strand carrying specific data information is assigned a corresponding primer. When the information of a particular DNA sequence needs to be extracted, PCR technology is employed to amplify the DNA strands in the DNA pool that contain the corresponding primer as the target strand. Post-PCR, a new DNA pool is formed, which includes multiple copies of the targeted DNA strands along with a handful of unrelated DNA strands.

$\emph{3)}$ DNA Sequencing: In this newly formed DNA pool, targeted information can be sequenced from the replicated copies of the target strand.

The process of synthesis, amplification, and sequencing, conceptualized as the DNA channel, can result in insertions, deletions, and substitutions of nucleotides in the DNA strand. Importantly, the model is designed considering the utilization of multiple PCR amplification copies for data decoding.

The structure of the channel model is depicted in Fig.~\ref{fig.propor1}. This layer simulates the three primary types of errors encountered in DNA: insertion (where extra nucleotides are added to the DNA sequence), deletion (where nucleotides are absent from their expected sequence positions), and substitution (where incorrect nucleotides replace the expected ones in the sequence). The outcomes from this channel layer are denoted as $\{\pmb{\hat{z}}^1,\pmb{\hat{z}}^2,..., \pmb{\hat{z}}^v\}$, with $v$ representing the aggregate number of nucleotide sequences intended for decoding. In our efforts to train and evaluate the resistance of the system to interference, we emulate the DNA storage channel. Drawing from the findings of \cite{2020HEDGES}, within a standard DNA storage workflow, the respective error percentages for nucleotide insertion, deletion, and substitution stood at $17\%$, $40\%$, and $43\%$. It's pertinent to note that these percentages account for errors emerging during various phases, including synthesis, handling of samples, preparation for storage, and the sequencing process itself. In our emulation, we make the assumption that substitution, insertion, and deletion errors are not only mutually exclusive but also uniformly random across all nucleotides. We designate the overall error likelihood as $\gamma$. Using the error model detailed above, we systematically introduce insertion, deletion, and substitution errors to the sequence $\pmb{z}$, producing $\pmb{\hat{z}}$. This can be mathematically expressed as:
\begin{align}\label{eq.cha}
\pmb{\hat{z}}=\mathfrak{S}(\mathfrak{D}(\mathfrak{I}(\pmb{z}, 0.17 \gamma), 0.4 \gamma), 0.43 \gamma),
\end{align}
wherein, $\mathfrak{S}(a,b),\mathfrak{D}(a,b)$, and $\mathfrak{I}(a,b)$ respectively represent processes where each component in $a$ has a $b$ chance of having a nucleotide inserted, deleted, or substituted with an arbitrary nucleotide.

\begin{figure}[htbp]
  \centering
    \includegraphics[width=0.5\textwidth]{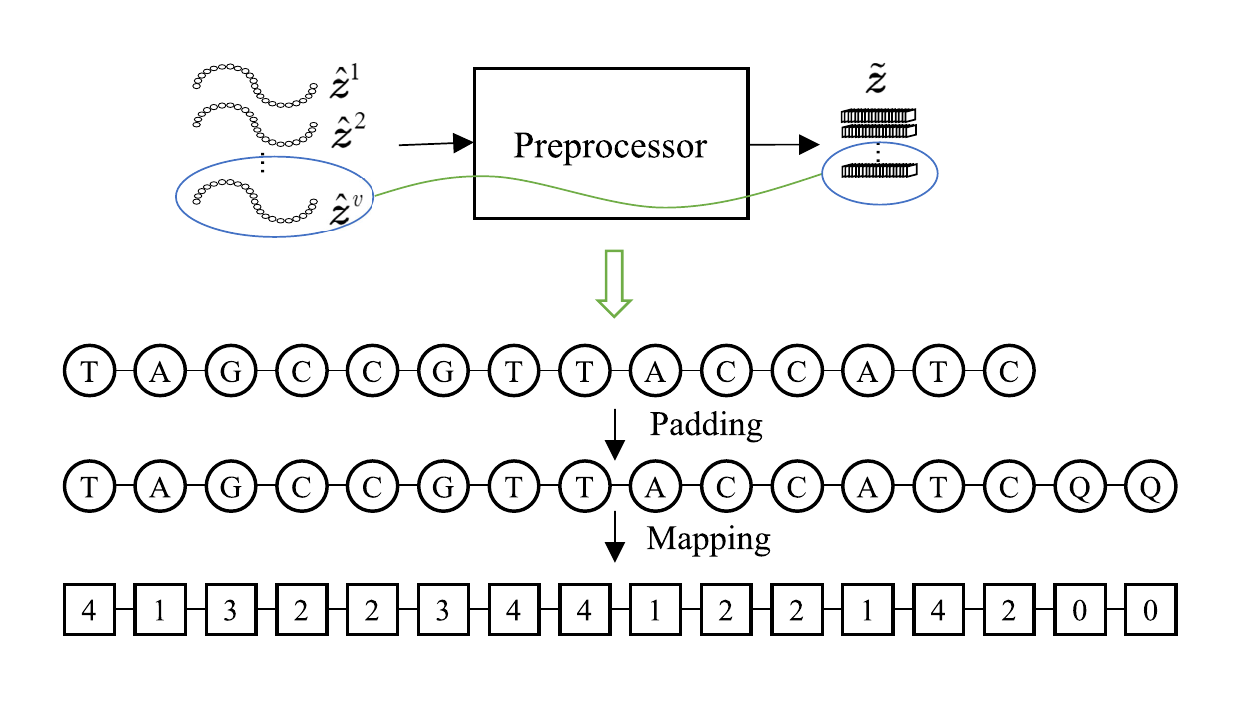}
    \caption{The structure of preprocessor model.}
\label{fig.propor2}
\end{figure}

As the number of insertions and deletions is typically unequal, the nucleotide sequences $\{\pmb{\hat{z}}^1,\pmb{\hat{z}}^2,..., \pmb{\hat{z}}^v\}$ can vary in length. To simplify the neural network computation, all sequences need to be processed to achieve uniform length and mapping to digital data. As shown in Fig.~\ref{fig.systemmodel}, we modularize this process into a preprocessor.

The architecture of the preprocessor is presented in Fig.~\ref{fig.propor2}.
Within the preprocessing layer, the size of the nucleotide sequence is normalized to $K$, with the length of $\{\pmb{\hat{z}}^1,\pmb{\hat{z}}^2,..., \pmb{\hat{z}}^v\}$ serving as the basis for determining the value of $K$. The principal aim here is to preserve the integrity of the original sequence data, padding shorter sequences with the symbol $Q$. It is imperative to maintain $K$ within reasonable limits to avert the inclusion of extraneous information. In general, the chosen value for $K$ is marginally greater than $k$, a process referred to as padding.
Subsequent to this, $\{Q,A,C,G,T\}$ are assigned numerical representations as $\{0,1,2,3,4\}$, respectively. All the processed sequences $\{\pmb{\hat{z}}^1,\pmb{\hat{z}}^2,..., \pmb{\hat{z}}^v\}$ are then amalgamated into $\pmb{\Tilde{z}}\in\mathbb{Z}^{v\times K}$ before being forwarded to the decoder. This procedure can be succinctly depicted as:
\begin{align}\label{eq.pro}
\pmb{\tilde{z}}=[\mathfrak{P}(\pmb{\hat{z}}^1), \mathfrak{P}(\pmb{\hat{z}^2}),..., \mathfrak{P}(\pmb{\hat{z}}^v)] \in\mathbb{Z}^{v \times K},
\end{align}
Here, $\mathfrak{P}(a)$ signifies the amalgamated padding and mapping operation.

The aforementioned channel and preprocessor modules, devoid of any learnable parameters, can be incorporated into the neural network architecture as non-training layers.

\subsection{Decoder}
\label{sec.Decoder}

The decoder translates $\pmb{\Tilde{z}}$ into a reconstruction of the initial image $\pmb{\hat{x}}$ via the function $g_\phi:\mathbb{Z}^{v\times K}\xrightarrow{} \mathbb{R}^{H\times W\times C} $, where $\phi$ signifies the parameter set of the joint source-channel decoder. The decoding procedure can be represented as:

\begin{align}\label{eq.dec}
\pmb{\hat{x}}=g_{\phi}(\pmb{\tilde{z}})\in\mathbb{R}^{H\times W\times C}.
\end{align}

The decoding architecture comprises a single-dimensional convolution layer and a CNN that is in reverse order to the compressor. The input of module, $\pmb{\tilde{z}}$, encompasses retransmission data. Initially, a one-dimensional convolution operation is executed to amalgamate the preliminary data, resulting in $\pmb{b}$, the dimensions of which should correspond with $\pmb{a}$. Thereafter, $\pmb{b}$ is inputted into a CNN, which is comprised of a two-dimensional transposed convolution layer, a Batch Normalization (BN) layer, and a PReLU activation function layer. The final output is normalized to the range of $[0, 255]$ by employing the sigmoid activation function layer and de-normalization layers. This network structure enables the high-dimensional reconstruction of the image, denoted as $\pmb{\hat{x}}$.


\section{DNA storage based end-to-end design}

The design of end-to-end DNA storage is intrinsically connected with a pivotal issue that necessitates resolution: biological constraints. These constraints encompass the homopolymer run-length restriction, also known as the Run-Length-Limited (RLL) constraint, and the GC-content constraint.

\begin{figure}[htbp]
  \centering
    \includegraphics[width=0.5\textwidth]{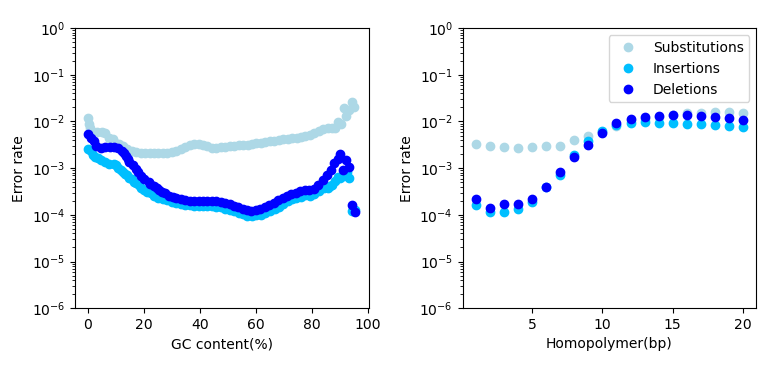}
    \caption{The error rate of Illumina sequencing as a function of GC-content and homopolymer run-length is represented\cite{2013Characterizing}.}
\label{fig.errorrate}
\end{figure}

The term 'homopolymer run-length' is utilized to describe the sequence length of consecutive, identical nucleotides within a DNA strand. Another important metric, the GC-content, represents the percentage composition of guanine (G) and cytosine (C) in a DNA sequence.
Studies have consistently shown that synthesis and sequencing inaccuracies predominantly arise due to extended homopolymer runs and extreme GC-contents, either high or low \cite{2013Characterizing, 2010Scalable}.
As a logical implication, it's advisable to enforce constraints on the length of homopolymer runs (termed the RLL constraint) and maintain the GC-content within an acceptable bandwidth (defined as the GC-content constraint).
For further clarity, Fig.~\ref{fig.errorrate} provides a comprehensive illustration. Ross et al pointed out that DNA sequences characterized by homopolymer run-lengths surpassing 5 nucleotides tend to manifest amplified deletion and insertion error rates. Additionally, sequences showcasing a GC-content above $70\%$ or below $40\%$ are more prone to base insertion and deletion errors during processes like Illumina sequencing. The empirical focus has generally been on constraining the GC-content around the $50\%$ mark \cite{2017DNA,2018A}.
Given these findings, our proposed model is tailored to ensure a minimized homopolymer run-length, while keeping the GC-content within the bounds of $45\%-55\%$.

Section \ref{sec.Loss function} delves into the optimization challenge focused on aligning nucleotide sequences $\pmb{z}$ with inherent biological restrictions, all the while optimizing the fidelity of the regenerated image $\pmb{\hat{x}}$.
The ensuing details concerning the end-to-end training regimen are exhaustively articulated in Section \ref{sec.training}.

\subsection{Optimization Problem}
\label{sec.Loss function}

We structure our optimization goal, encompassing the Reconstruction Quality (RQ) and Biological Constraints (BC) sections.

$\emph{1) RQ segment:}$ This section elaborates on the performance of the DJSCC-DNA methodology. We employ Mean Squared Error (MSE) to quantify the average deviation between the reconstructed image $\pmb{\hat{x}}$ and the original image $\pmb{x}$:
\begin{align}\label{eq.LRQ}
\mathcal{L}_{RQ}=\frac{1}{n} \sum_{i=1}^{n} (\pmb{x}_i,\pmb{\hat{x}}_i)^2.
\end{align}

$\emph{2) BC segment:}$ The BC component is intended to constrain nucleotide sequences $\pmb{z}$ to adhere to biological limitations as much as possible. Let us assume that a metric $S(\pmb{z})$ evaluates the GC-content and homopolymer run-length of sequence $\pmb{z}$, and a computable $S^*$ symbolizes its target values. The $\mathcal{L}_{BC}$ is the deviation between $S(\pmb{z})$ and $S^*$:
\begin{align}\label{eq.L_bc}
\mathcal{L}_{BC}=\frac{1}{m} \sum_{i=1}^{m} (S_i(\pmb{z}), S^*)^2,
\end{align}
where $m$ is the size of $S(\pmb{z})$ and $S_i(\pmb{z})$ donates the $i$-th element in $S(\pmb{z})$.

For sequence $\pmb{z}\in\mathbb{Z}^k$, the values $\{0,1,2,3\}$ are representative of the four nucleobases $\{\mathrm{A,C,G,T}\}$ respectively. For a DNA sequence depicted by $\pmb{z}\in\mathbb{Z}^k$, each value in the set $\{0,1,2,3\}$ corresponds to one of the four nucleotide $\{\mathrm{A,C,G,T}\}$. The occurrence frequency of each nucleotide is indicated by the vector $\{p_1, p_2, p_3, p_4\}$, where $p_i$ symbolizes the proportion of the sequence composed of the $i$-th nucleobase. The $S(\pmb{z})$ should include the GC content and homopolymer run-length measurements, which can be expressed as:
\begin{align}\label{eq.Sth}
S(\pmb{z})&=(\mathcal{T}(\pmb{z}), \mathcal{H}(\pmb{z}) ) ,
\end{align}
where $\mathcal{T}(\pmb{z})$ denotes the GC-content measurement index, and $\mathcal{H}(\pmb{z})$ signifies the homopolymer run-length measurement index. $S^*=(\mathcal{T}^*, \mathcal{H}^*)$ are their corresponding target values.

The GC content of the sequence is tantamount to the ratio of 1 and 2 in the vector $\pmb{z}$. Initially, it should be stressed that having a relatively equilibrated probability of occurrence for each base in the DNA sequence is desirable, as this assures a higher information content in the sequence. When the probability of a certain base is imbalanced, it can trigger a reduction in the amount of unique information conveyed by the DNA sequence. This can obfuscate different parts of the sequence and potentially affect sequence stability. Consequently, we presume that the proportions of A and T, and G and C are equal. The mean of the vector value can be used as a GC-content measurement. Even if the quantities of A-T or G-C are unequal, the goal of achieving a $50\%$ GC-content is feasible with a fluctuation of $5\%$.

To minimize the homopolymer run-length in a nucleotide sequence, it's advantageous to promote a uniform distribution of the four nucleotides in sequences. Take, for example, the sequences $-\mathrm{AAAAAAGGGGGGTTTTCCCCCCTT}-$ and $-\mathrm{ACAGAATAGGCTTCGGTCCACGTT}-$. While both maintain identical values for $p_1, p_2, p_3, p_4$, the latter exemplifies a more balanced sequence. The underlying principle is to minimize the prevalence of recurring nucleotides in a specific DNA sequence segment, thus ensuring a consistent nucleobase spread throughout. This perspective frames the DNA sequence in terms of overlapping segments, wherein the variance of each segment can be analyzed to determine its nucleobase equilibrium. In alignment with our prior optimization goal of realizing an equal nucleotide distribution ($p_1 = p_2 = p_3 = p_4$), a pronounced variance signifies a predominance of A-T bases in the segment (implying $p_1 + p_4$ is elevated), whereas a diminished variance indicates a surplus of G-C bases ($p_2 + p_3$ is heightened). By modulating the variance of these segments to fit within a defined range, it becomes feasible to regulate the A-T or G-C base pair concentration within any segment. Such regulation not only curtails the homopolymer run-length but also ensures a more harmonized nucleobase dispersal in the DNA sequence.

In summary, the nucleotide sequence $\pmb{z}$, drawn from the set $\mathbf{Z^k}$, is subdivided into $m$ segments, each spanning a length of $d$. These segments are structured with an interspacing equivalent to half their length ($d/2$). The initial positions of these segments are denoted by:
\begin{align}
    \label{eq.mm}
    \mathcal{M}=[1,1+d/2,1+d,...,1+d/2*(m-1)],
\end{align}
where $d$ is a determinant signifying the rigorousness of the evaluation and is stipulated to be even. The cumulative segment count is denoted as $m =\lceil2k / d\rceil-1$.
For each segment, the central tendency and dispersion are articulated as:
\begin{subequations}\label{eq.th}
\begin{align}
   \mathcal{T}_i &= \frac{1}{d}\sum_{t=\mathcal{M}_i}^{\mathcal{M}_i+d} \pmb{z}_t , \\
   \mathcal{H}_i &= \frac{1}{d-1}\sum_{t=\mathcal{M}_i}^{\mathcal{M}_i+d} ({\pmb{z}_t-\mathcal{T}_i})^2.
\end{align}
\end{subequations}
Here, $\pmb{z}_t$ represents the $t$-th nucleotide within $\pmb{z}$ and $\mathcal{M}_i$ denotes the $i$-th constituent of $\mathcal{M}$. The index $S(\pmb{z})$ assimilates $\mathcal{T}(\pmb{z})$ and $\mathcal{H}(\pmb{z})$, and can be delineated as:
\begin{subequations}\label{eq.Sz}
\begin{align}
   \mathcal{T}(\pmb{z}) &= [\mathcal{T}_1, \mathcal{T}_2,...,\mathcal{T}_m ],\\
   \mathcal{H}(\pmb{z}) &= [\mathcal{H}_1, \mathcal{H}_2,...,\mathcal{H}_m ]
\end{align}
\end{subequations}

Given an assumed GC-content of $50\%$, where $p_1=p_2=p_3=p_4=0.25$, the benchmark values of $S^*$ for a sequence that aligns with both constraints are approximately:

\begin{subequations}\label{eq.Sxing}
\begin{align} 
  \mathcal{T}^* &= 0*p_1+1*p_2+2*p_3+3*p_4  \\
  &= 1.5, \\
  \mathcal{H}^* &= 0.5^2*(p_2+p_3)+1.5^2*(p_1+p_4), \\
  &=1.15,\\
   S^*&=(\mathcal{T}^*, \mathcal{H}^*)=(1.5,1.25).
\end{align}
\end{subequations}

\begin{figure}[htbp]
  \centering
    \includegraphics[width=0.5\textwidth]{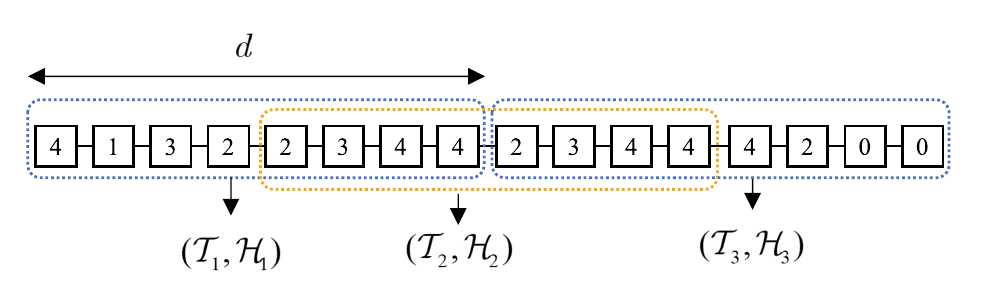}
    \caption{An example of $S(\pmb{z})$ with $d=8$ and $m=3$}
\label{fig.lossbc}
\end{figure}

An illustration of $S(\pmb{z})$ is provided in Fig.~\ref{fig.lossbc}. This representation involves segmenting the nucleotide sequence into overlapping areas of length $d$, followed by the calculation of the mean and variance for each area.

Subsequently, incorporate Eq. (\ref{eq.Sz}) and Eq. (\ref{eq.Sxing}) into Eq. (\ref{eq.L_bc}) to obtain the sum of the two parts, and magnify the $\mathrm{part~2}$ by a factor of 10. We express $\mathcal{L}_{BC}$ as:
\begin{align}\label{eq.LBC}
\mathcal{L}_{BC}= \underbrace{ \frac{1}{m} \sum_{i=1}^{m} (\mathcal{T}_i,\mathcal{T}^*)^2}_{\mathrm{part 1}}+ \underbrace{  \frac{10}{m} \sum_{i=1}^{m} (\mathcal{H}_i,\mathcal{H}^*)^2 }_{\mathrm{part 2}},
\end{align}
here, $\mathrm{part~1}$ represents the GC-content constraint, while $\mathrm{part~2}$ signifies the RLL constraint. Because of the small value of $\mathrm{part ~2}$, this constraint could be overlooked during the backpropagation phase of the neural network. To circumvent this, the constraint is amplified by a factor of 10, thereby magnifying its influence on the network optimization process.

Based on these considerations, the objective function of  the optimization problem can be framed as:

\begin{subequations}\label{eq.min}
\begin{align}
   \underset{\theta,\phi}\min ~&  \mathcal{L}_{RQ}+\alpha \mathcal{L}_{BC} ,  \\
   \text{s.t.} ~&(\ref{eq.enc}),(\ref{eq.cha}),(\ref{eq.pro})  ~\text{and}~ (\ref{eq.dec}),   \\
    &\alpha > 0 ,
\end{align}
\end{subequations}
where $\alpha$ is an tunable parameter.

\subsection{End-to-End Training}
\label{sec.training}

The schematic diagrams of the encoder and decoder used in our methodology are furnished in Tab.\ref{tab.Structureparameters}. Here, the tuple notation $(C,K,S)$ signifies a convolutional layer composed of $C$ filters, each with a kernel size $K$ and a stride $S$.

\begin{table}[tbp]
\centering
\caption{Structure parameters}
\begin{tabular}{l  l}
\toprule
Encoder                   & Decoder
 \\ \midrule
 Normalization     &   Conv1d (1,3,1), PReLU               \\
Conv2d (16,3,2), PReLU     &   T-Conv2d (32,3,1), PReLU          \\
Conv2d (32,3,1), BN, PReLU &   T-Conv2d (32,3,1), BN, PReLU        \\
Conv2d (32,3,1), BN, PReLU &  T-Conv2d (32,3,1), BN, PReLU       \\
Conv2d (32,3,1), BN, PReLU & T-Conv2d (16,4,2), BN, PReLU       \\
Conv2d (c,3,1)             &  T-Conv2d (3,4,2), Sigmoid          \\
$round~(\text{Sigmoid}\times 3)$    &   Denormalization  \\
\bottomrule
\end{tabular}
\label{tab.Structureparameters}
\end{table}

Within the encoder, the input image values, structured as $32\times 32\times 3$, are uniformly initialized within the interval $[0,1]$. The compression component is structured as follows: it comprises combined layers, which include two-dimensional convolution + PReLU, 3-th two-dimensional convolution + BN + PReLU, and an additional two-dimensional convolution. The final convolutional layer produces feature size $k=c\times 8\times 8$, with parameters set analogous to the methodologies in \cite{2019deep} and \cite{2018an}. The nucleotide pixel ratio is represented as $R=k/n$. We can modulate $R$ by varying the value of $c$.

For the decoder, the channel output samples, dimensioned as $v \times K$, are subject to a one-dimensional convolution to yield $k$ values. These values are subsequently supplied to the inverse CNN. This network is accountable for inverting the operations executed by the encoder, involving two-dimensional transpose convolutional layers, BN layers, and PReLU layers. The concluding two-dimensional transpose convolutional layer is linked to a Sigmoid activation function to constrain the output within the $[0,1]$ interval. Lastly, the values are uniformly rescaled to span the $[0,255]$ range.

The DNA channel and preprocessor are constituted as non-trainable layers. In light of the present state of DNA synthesis techniques, synthesizing long-chain DNA is both labor-intensive and inefficient. Therefore, we partition the output sequence of  the encoder into non-overlapping segments, each of length $s$, which are then recombined before being fed into the decoder.

The loss function is expressed as follows:
\begin{equation}
\begin{aligned} 
\label{eq.Lossfunction}
    \mathcal{L}(\theta,\phi)=&\frac{1}{n} \sum_{i=1}^{n} (\pmb{x}_i,\pmb{\hat{x}}_i)^2 +
    \\&\alpha \times [ (\frac{1}{m} \sum_{i=1}^{m} (\mathcal{T}_i,\mathcal{T}^*)^2+ \frac{10}{m} \sum_{i=1}^{m}(\mathcal{H}_i,\mathcal{H}^*)^2 ].
\end{aligned}
\end{equation}

\section{Numerical Results}
\label{sec.resultsbig}

This section begins with an examination of the influence of the parameter $\alpha$ on the model proposed in \ref{sec.partition}, followed by a bifurcation of the model into two specific variants. Subsequently, a detailed performance comparison and analysis of the introduced variants are provided in \ref{sec.results}.
\subsection{Simulation Setup }

The architecture as delineated in Section \ref{sec.training} is implemented using Pytorch. The Adaptive Moment Estimation (Adam) optimizer is utilized during the optimization process. The complete model undergoes training for 50,000 iterations, with an initial learning rate of $5\times 10^{-3}$, subsequently reduced to $5\times 10^{-4}$. The CIFAR-10 image dataset is employed for evaluating our scheme. Table \ref{tab.parameter} lists the particulars of the simulation parameters. The effectiveness of this approach is gauged by evaluating the peak signal-to-noise ratio (PSNR) and structural similarity index (SSIM).

\begin{table}[tbp]
\centering
\caption{Simulation parameters}
\begin{tabular}{l l}
\toprule
Parameters   &  Values
 \\ \midrule
$n=H\times W\times C$     & $3072=32\times 32\times 3$ \\
$R=k/n$ (nts/pixel) &  $[0.125,0.5,0.75,1,1.25,1.5]/3$ \\
$\gamma_{\mathrm{train}}$    &  $0.50\%$   \\
$v$    &  $2$    \\
$d$    &  $8$   \\
$s$     & $256$ \\
$\alpha$  &   $[75~(\text{Proposed-1}),175~(\text{Proposed-2})]$   \\

\bottomrule
\end{tabular}
\label{tab.parameter}
\end{table}

\subsection{Model partition}
\label{sec.partition}

Initially, an evaluation of the performance of the algorithm with respect to image quality and biological constraints for various $\alpha$ values is performed. $\alpha$ stands for the weightage assigned to the biological constraint term in the loss function given by Eq. (\ref{eq.Lossfunction}), thereby determining the constraint strength the biological aspects impose on the model during the generation of the nucleotide sequence by the encoder.

\begin{figure}[htbp]
  \centering
    \includegraphics[width=0.5\textwidth]{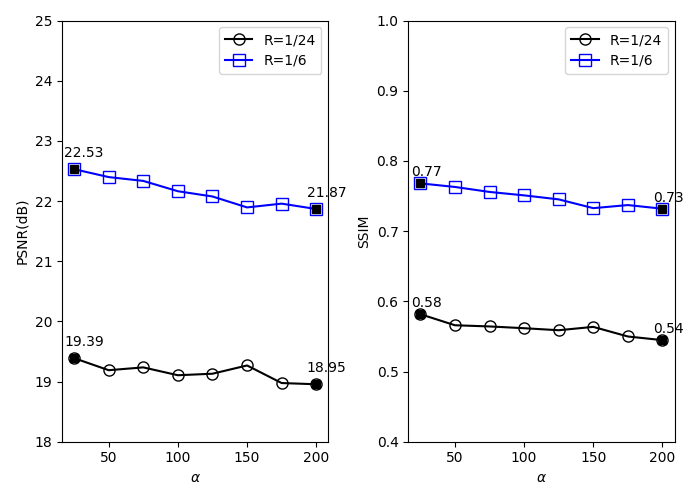}
    \caption{Performance of the proposed DJSCC-DNA with different $\alpha$ on CIFAR-10 test images.}
\label{fig.compara_perfor}
\end{figure}

Fig.~\ref{fig.compara_perfor} depicts PSNR and SSIM values for $R$ values of 1/24 and 1/6 as $\alpha$ ranges from 10 to 200. With an increase in $\alpha$, both the PSNR and SSIM between the original and reconstructed images decline, implying a reduction in the quality of the restored image. This is anticipated, as when the biological constraints receive a higher weightage in the loss function during optimization, the corresponding requirement for image quality restoration decreases. The blue and black lines in the graph represent $R$ values of 1/6 and 1/24, respectively, with maximum PSNR differences of 0.66 and 0.44. The maximum SSIM difference between the two lines is 0.04, a difference that is nearly indistinguishable to the unaided eye. Overall, the influence of the $\alpha$ value on image restoration quality appears to be within a tolerable range.

\begin{figure}[htbp]
  \centering
    \includegraphics[width=0.5\textwidth]{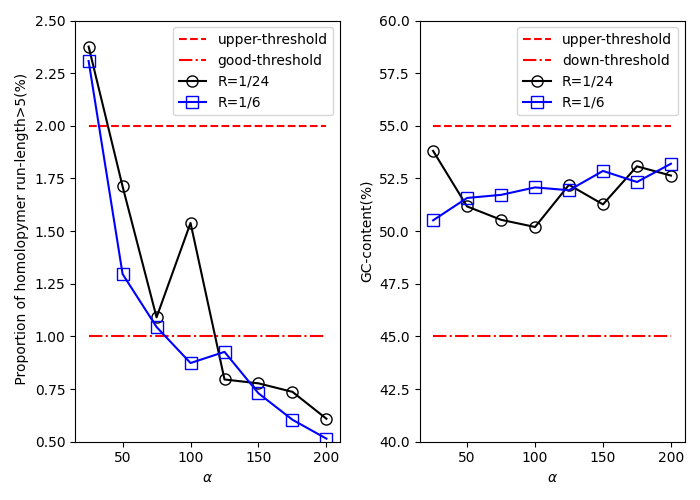}
    \caption{Biological constraints of proposed DJSCC-DNA with different $\alpha$ on CIFAR-10 test images.}
\label{fig.compara_constr}
\end{figure}

Fig.~\ref{fig.compara_constr} presents data statistics related to biological constraints of the nucleotide sequence generated by the encoder, retaining the same parameter settings as in Fig.~\ref{fig.compara_perfor}. On the left, the mean proportion of nucleotides with a homopolymer length greater than $5$, relative to the total count of nucleotides in the produced nucleotide sequence, is computed. The right side displays the proportion of GC content. As an example, for the sequence $-\mathrm{AAAAAAGATGTG}-$, the proportion of homopolymer run-length greater than $5$ stands at $50\%$ with the GC-content at $0.25\%$.

Assuming the satisfaction of biological constraints, the proportion of homopolymer run length greater than $5$ should not exceed $2\%$, with less than $1\%$ being more desirable. Further, the GC-content is ideally within the $45-55\%$ bracket. These thresholds are marked as red dotted lines in the figure. It can be observed that with an increase in $\alpha$, the proportion of bases with a polymer length over $5$ decreases, falling below $1\%$ when $\alpha$ surpasses $100$. Moreover, the GC content of the sequence consistently resides within the desired range of $45-55\%$. These findings suggest that our crafted loss function operates in accordance with our assumptions.

Building on the selected values of $\alpha$, we carried out exhaustive testing and analysis of the performance of our two proposed methods, referred to as Proposed-1 ($\alpha=75$) and Proposed-2 ($\alpha=175$).

\subsection{Results}
\label{sec.results}

In our paper, we initially compared the conventional neural network-based technique with our proposed scheme on a given test set. The conventional technique, VAEU-QC, utilizes the Variational Auto-Encoder (VAE) method for data compression and quaternion coding (QC) for base nucleobase coding \cite{2021generative}. The neural network within the VAEU extracts latent variables from a singular-channel image of dimensions $32\times 32$. These variables are subsequently transmitted through a fully connected layer to the deconvolution network. The network structure parameters were adjusted as recommended in \cite{2021generative}. During the CIFER-10 image dataset evaluation, the three channels were concurrently encoded and spliced. It is important to note that error correction coding was not employed in this scheme. We approximated the impact of error correction on the comparative method by setting the channel noise $\gamma_{\mathrm{test}}$ to $0$, implying an almost flawless error correction scheme within the comparative method.

\begin{figure}[htbp]
  \centering
    \includegraphics[width=0.5\textwidth]{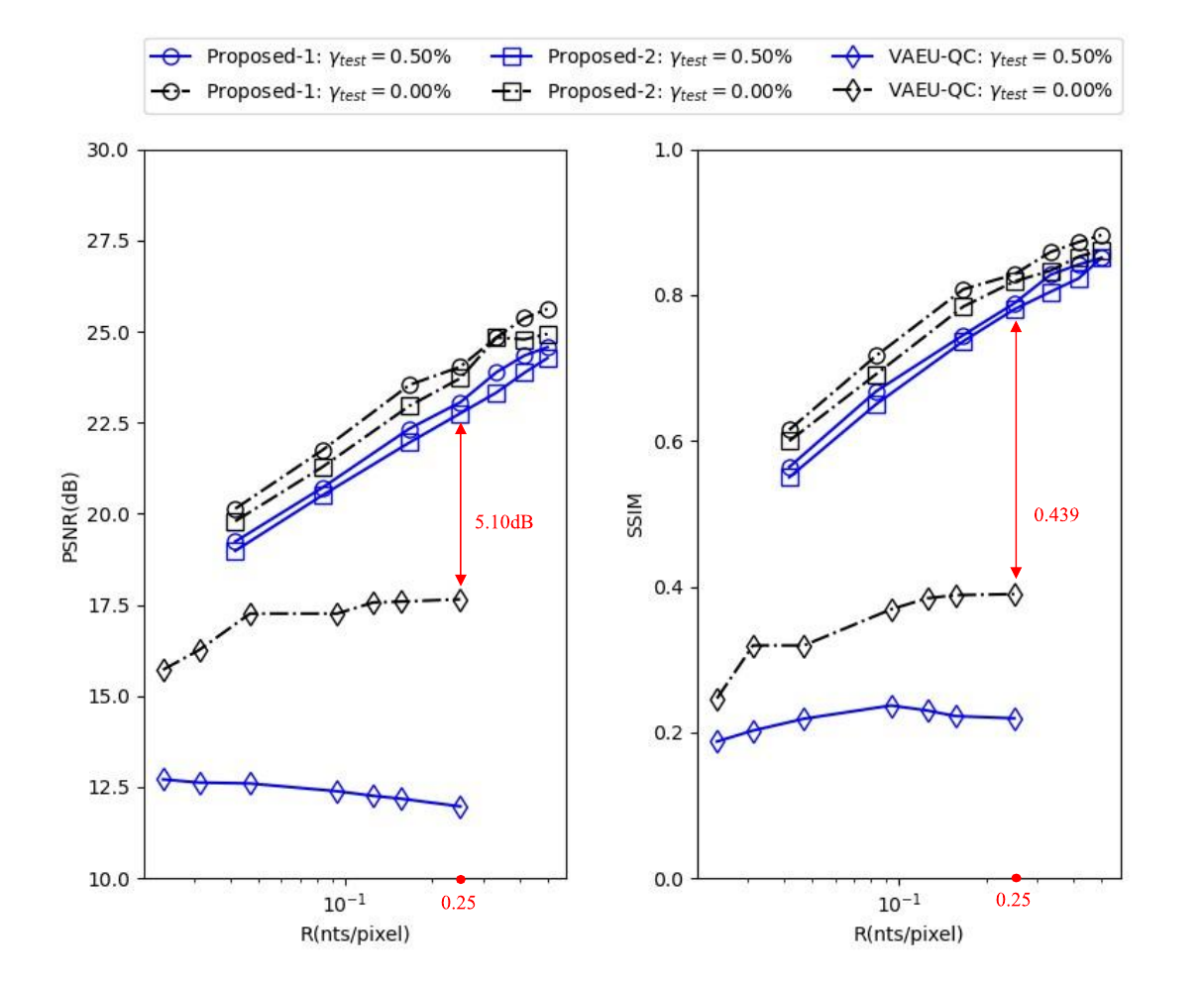}
    \caption{Performance of VAEU-QC and proposed DJSCC-DNA on CIFAR-10 test images.}
\label{fig.compara_vae}
\end{figure}

The different schemes' performance is represented in Fig.~\ref{fig.compara_vae}. The proposed DJSCC-DNA scheme exhibits superior performance in image reconstruction. Specifically, as the nucleotide pixel ratio $R$ grows, the PSNR of the proposed methods with $\gamma_{\mathrm{test}}=0.50\%$ shows an ascending trend, predominantly ranging between $18$ $\mathrm{dB}$ and $23$ $\mathrm{dB}$, while for the VAEU-QC scheme with $\gamma_{\mathrm{test}}=0$, it falls between $15.5$ $\mathrm{dB}$ and $17.5$ $\mathrm{dB}$.

Moreover, the SSIM of the proposed DJSCC-DNA scheme with $\gamma_{\mathrm{test}}=0.50\%$ mostly varies between $0.55 $ and $0.85$, while it is consistently between $0.25$ and $0.40$ for the VAEU-QC scheme with $\gamma_{\mathrm{test}}=0$. Further, when $R=0.25$, the PSNR and SSIM of the proposed scheme are approximately $5.10$ $\mathrm{dB}$ and $0.439$ higher than the VAEU-QC scheme, respectively. The proposed algorithm demonstrates improved performance over the VAEU-QC scheme in terms of PSNR and SSIM. While the VAEU-QC scheme is capable of efficiently storing images with a high proportion of single colors, such as those in the MNIST image set referenced in \cite{2021generative}, it may not be suitable for storing more complex color images. Contrarily, the proposed scheme effectively addresses such complex images.

As illustrated in Fig.~\ref{fig.compara_vae}, the PSNR and SSIM of the VAEU-QC scheme, with a noise parameter of $\gamma_{\mathrm{test}}=0.50\%$, hold steady at approximately $13$ $\mathrm{dB}$ and $0.2$, respectively. The lack of an error correction scheme in this scenario, coupled with the significant influence of DNA channel noise, is responsible for this trend. Conversely, in the proposed methodology, the PSNR and SSIM values at the same noise level are minimally diminished (by approximately $0.5$ $\mathrm{dB}$ and $0.5$), compared to their counterparts at $\gamma_{\mathrm{test}}=0$, with this difference narrowing as the $R$ value ascends. This finding alludes to the ability of the proposed methodology to rectify errors within a JSCC framework, also established on a neural network foundation. Additionally, as $R$ increases, the resilience of the system to errors rises, validating the proficiency of the proposed approach in mitigating system errors.

\begin{figure*}[htbp]
  \centering
    \includegraphics[width=0.85\textwidth]{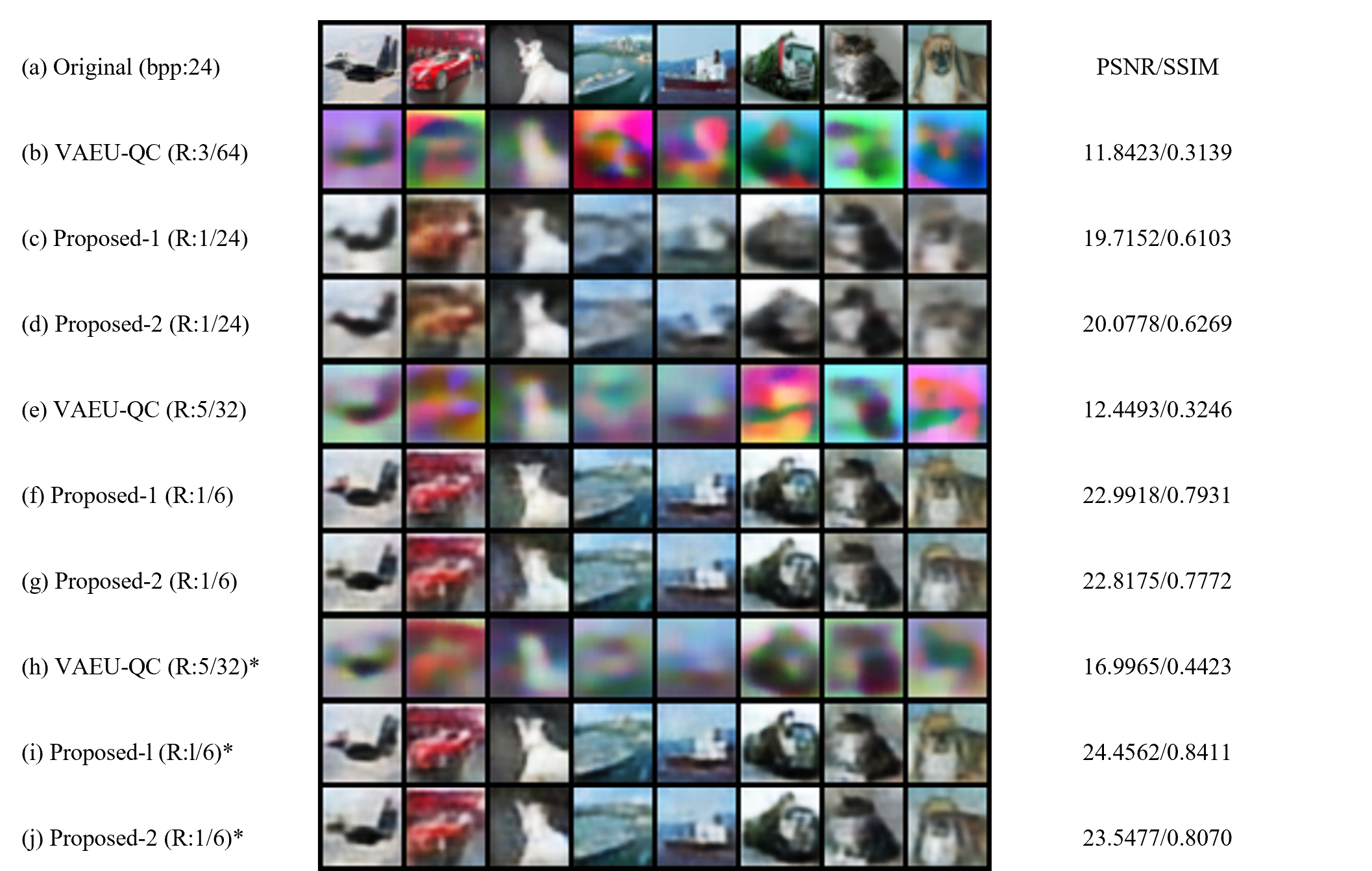}
    \caption{Performance of VAEU-QC scheme and proposed DJSCC-DNA scheme on CIFAR-10 test images: (b-g) are tested in $\gamma=0.50\%$; (h-j) marked with '*' are tested in $\gamma=0$}
\label{fig.compara_small}
\end{figure*}

Fig.~\ref{fig.compara_small} visually compares the methodologies, with the left segment representing the training methodologies. Here, the hexagon symbolizes $\gamma_{\mathrm{test}}=0$, while the unmarked graphic designates $\gamma_{\mathrm{test}}=0.50\%$. The right segment presents the average test PSNR and SSIM for each methodology. All results depicted are extrapolated from the data points displayed in Fig.~\ref{fig.compara_vae}, facilitating an intuitive comprehension of our prior conclusions.

\begin{figure}[htbp]
  \centering
    \includegraphics[width=0.5\textwidth]{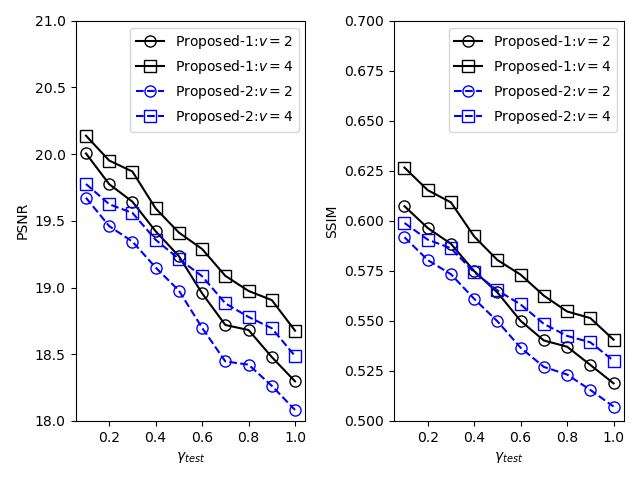}
    \caption{Performance of the proposed DJSCC-DNA scheme ($R=1/24$) in different $v$ on CIFAR-10 test images.}
\label{fig.Results_v}
\end{figure}

The complexity of the model is evaluated through an analysis of four critical network parameters: model parameter quantity, model storage content, multiply-adds (MAdd), and floating-point operations (FLOPs), as shown in Tab.\ref{tab.analysis}. These metrics are routinely used to quantify the computational complexity and storage demands of a neural network model.

$\emph{1)}$ Encoder: CV (32,1,3), BN, ReLU; CV (64,32,3), BN, ReLU; CV (63,128,3); FC (32).

$\emph{2)}$ Decoder: FC (1152), DC (128,64), BN, ReLU; DC (64,32), BN, ReLU, DC (32,1), Sigmoid.

The model parameter quantity is a reference to the total count of trainable parameters within the model, which is intrinsically tied to the capacity of the model and learning potential. The storage content signifies the storage space required to accommodate the parameters of the model. The proposed model, which is optimized for processing 3-channel images, markedly surpasses the VAE model in both parameter and memory requirements.

MAdd and FLOPs are common metrics reflecting the computational complexity of the model. MAdd denotes the total count of multiply-add operations within the model, while FLOPs represent the total number of floating-point operations. The proposed model, designed to augment the extraction of image features, possesses deeper layers and higher-dimensional computations, leading to elevated MAdd values and decreased FLOPs in comparison to the VAE model.

\begin{table*}[tbp]
\centering
\caption{Model analysis}
\begin{tabular}{l l  l  l   l   l  l }
\toprule
Schemes  & Input size & R (nt/pixel)  & Total params & Total memory (MB)  &  Total MAdd (M)  & Total Flops (M)
 \\ \midrule
Proposed &(32,32,3)    &1/8  &   54921  & 0.24  &  7.47M  &  1.76M              \\

VAE &(32,32,1)    &1/8     &   297633  & 0.25  &  6.95M  &  1.82M              \\

\bottomrule
\end{tabular}
\label{tab.analysis}
\end{table*}

Within our model, the quantity of retransmission sequences, $v$, exerts a pivotal influence on the actual performance of the neural network. We conduct training of the proposed methodologies on $v=2,4$ and carry out performance tests under varying $\gamma_{\mathrm{test}}$ values as shown in Fig.~\ref{fig.Results_v}. Here, a circle denotes a $v$ value of $2$, while a square signifies $4$. Maintaining $v$ constant while increasing $\gamma_{\mathrm{test}}$ induces higher channel noise levels, which subsequently trigger declines in both PSNR and SSIM. Notably, when $\gamma_{\mathrm{test}}$ spans from $0.1\%$ to $1\%$, the average difference in PSNR and SSIM is approximately $1.5$ $\mathrm{dB}$ and $0.09$, respectively, suggesting a critical role of $\gamma_{\mathrm{test}}$ in determining the quality of image restoration.

Interestingly, we notice that as $v$ amplifies, both PSNR and SSIM ascend, a trend that becomes more pronounced with rising $\gamma_{\mathrm{test}}$ values. More specifically, when $\gamma_{\mathrm{test}}$ equals $0.1\%$ and $1\%$, PSNR increases by around $0.1$ $\mathrm{dB}$ and $0.5$ $\mathrm{dB}$ respectively, while SSIM amplifies by roughly $0.01$ and $0.025$. These outcomes underline the fact that increasing $v$ can enhance the quality of image restoration, particularly in environments with adverse channel conditions.

Given the considerable time complexity involved in channel simulation, we have opted to default to $v=2$. The primary goal here is to validate the practicability of the model.

\begin{figure}[htbp]
  \centering
    \includegraphics[width=0.5\textwidth]{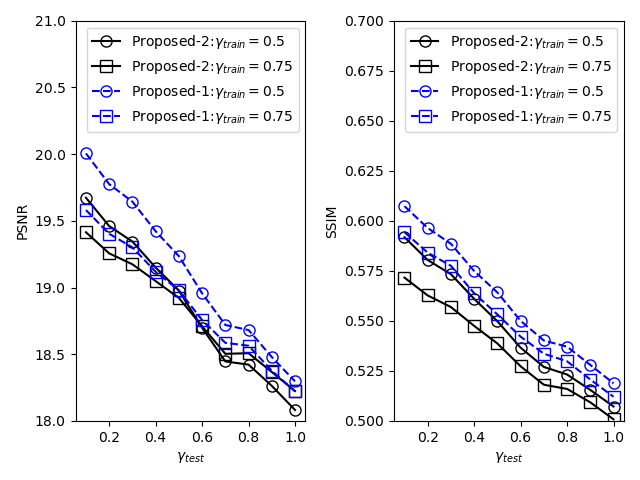}
\caption{Performance of the proposed DJSCC-DNA scheme ($R=1/24$) in different $\gamma_{\mathrm{train}}$ on CIFAR-10 test images.}
\label{fig.Results_y}
\end{figure}

Experiments were conducted to assess the performance of the proposed algorithm under various $\gamma_{\mathrm{train}}$ values, as exhibited in Fig.~\ref{fig.Results_y}. In this figure, the circle denotes $\gamma_{\mathrm{train}}=0.5\%$, while the square represents $\gamma_{\mathrm{train}}=0.75\%$. The findings suggest that $\gamma_{\mathrm{test}}$ lies within the range of $0.1\%$ and $1\%$, and in the majority of cases, superior algorithm performance is achieved when $\gamma_{\mathrm{train}}=0.5\%$. Furthermore, as $\gamma_{\mathrm{test}}$ incrementally increases, the performance disparity between $\gamma_{\mathrm{train}}=0.75\%$ and $\gamma_{\mathrm{train}}=0.5\%$ substantially diminishes. These results corroborate that, even in the absence of precise knowledge of specific channel parameters, a more secure $\gamma_{\mathrm{train}}$ can be selected for training by considering a rough estimation of channel noise. Hence, the proposed model retains a degree of universality for differing channel parameters and shows potential for practical application.

\begin{figure}[htbp]
  \centering
    \includegraphics[width=0.5\textwidth]{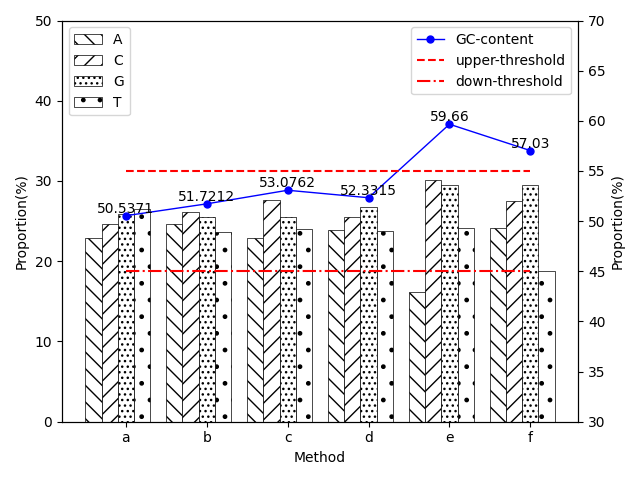}
    \caption{The proportion of $\mathrm{A,G,C,T}$ of CNN+M and proposed DJSCC-DNA schemes ($\gamma_{\mathrm{train}}=\gamma_{\mathrm{test}}=0.50\%$) on CIFAR-10 test images: From left to right and top to bottom, the corresponding schemes are (a) Proposed-1 in $R=1/24$, (b) Proposed-1 in $R=1/6$, (c) Proposed-2 in $R=1/24$, (d) Proposed-2 in $R=1/6$, (e) CNN-M in $R=1/24$, (f) CNN-M in $R=1/6$}
\label{fig.Results_gc}
\end{figure}

\begin{figure}[htbp]
  \centering   \includegraphics[width=0.5\textwidth]{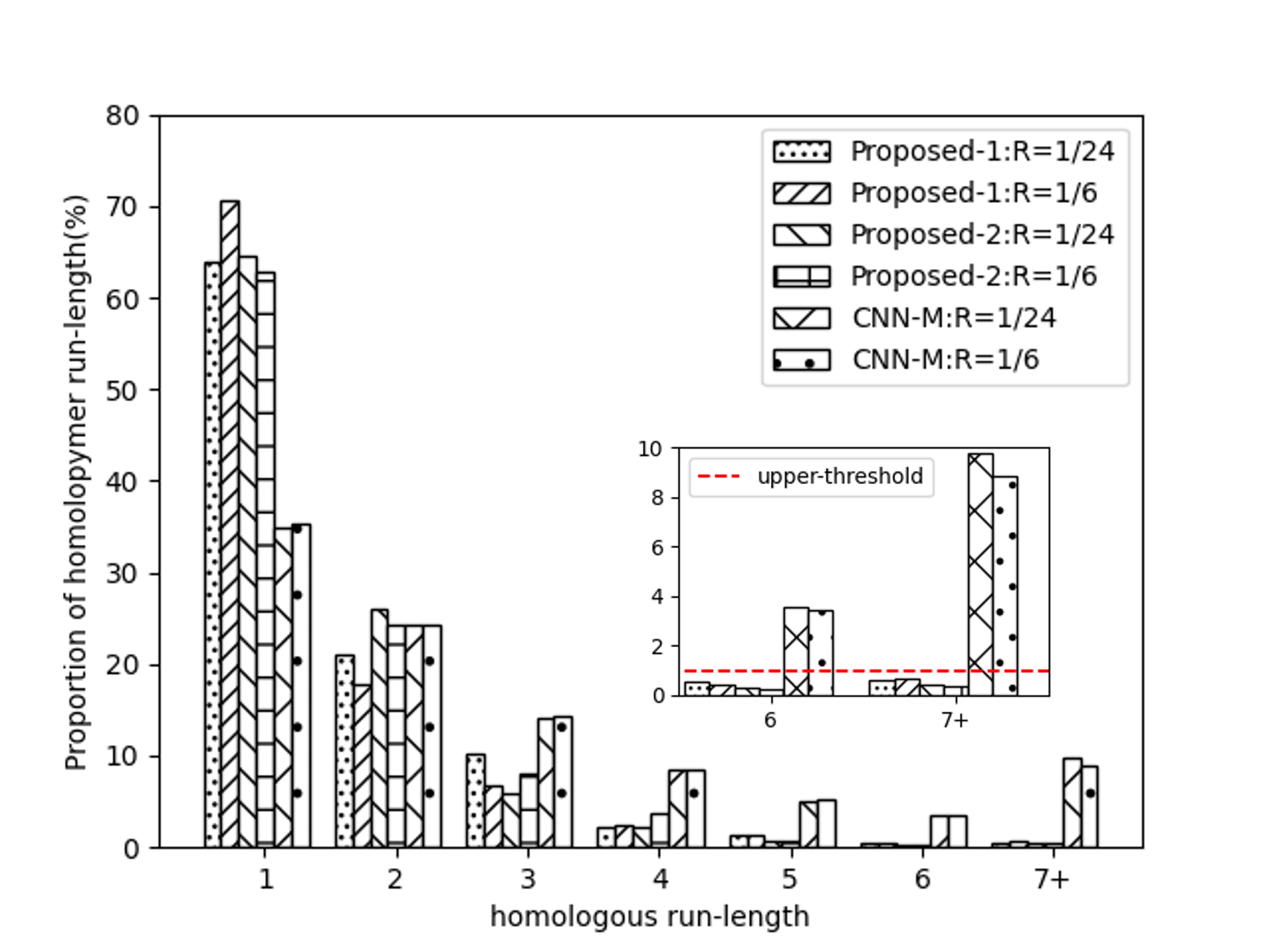}
    \caption{The proportion of homologous length of CNN+M and proposed DJSCC-DNA schemes ($\gamma_{\mathrm{train}}=\gamma_{\mathrm{test}}=0.50\%$) on CIFAR-10 test images.}
\label{fig.Results_rtt}
\end{figure}

In Section \ref{sec.Loss function}, the loss function was restructured to ensure the nucleotide sequence adheres to biological constraints. To substantiate the efficacy of this function, we scrutinize the biological characteristics of the nucleotide sequences obtained through multiple methodologies, as depicted in Fig.~\ref{fig.Results_gc} and Fig.~\ref{fig.Results_rtt}. We juxtapose the performance of the proposed methodologies, which utilize the newly designed loss function, with that of a conventional CNN model (CNN-M) that employs the traditional loss function.

Fig.~\ref{fig.Results_gc} enumerates the proportions of the four nucleotide types $\{\mathrm{A,C,G,T}\}$ and the GC-content in the nucleotide sequences. The horizontal axis delineates different methodologies, as explained in the figure caption. Here, schemes a-d are the schemes proposed in this study, while schemes e-f refer to the CNN-M schemes. The red dotted line demarcates the previously mentioned GC content threshold. As discernible from Fig.~\ref{fig.Results_gc}, the GC content of the nucleotide sequences produced by the proposed methodology falls within the range designated by the red dotted line, while that of CNN-M exceeds the upper threshold. Additionally, the $\{\mathrm{A,G,C,T}\}$ content of the sequences generated by the proposed methodology is more balanced compared to those produced by CNN-M, which is preferable. These findings illustrate the efficacy of the novel loss function, crafted in this study, in regulating GC content within the anticipated range.

Fig.~\ref{fig.Results_rtt} demonstrates the proportion of homologous lengths every $s$ nt, where the horizontal axis represents the homologous run length in nucleotide sequences. The inset of the figure offers a magnified view of the proportion when the homologous run length exceeds 5, and the red dotted line indicates the upper threshold. It is evident that when the length of the homopolymer surpasses $3$, the proportion generated by the CNN-M schemes is significantly higher than that of the proposed methodologies. This suggests that the nucleotide sequences produced by CNN-M have more and longer homopolymer run lengths, which markedly violates biological constraints. For instance, the proportion of homopolymer run length exceeding 5 is $13.31\%$ and $12.29\%$ for CNN-M, respectively. Conversely, with the constraints imposed by the loss function proposed in this research, the proportion of homopolymer run length exceeding 5 does not surpass $1.09\%$, $1.04\%$, $0.74\%$, and $0.60\%$, respectively. These findings underline the positive influence of the newly developed loss function on conforming nucleotide sequences to biological constraints.

\begin{figure}[htbp]
  \centering
    \includegraphics[width=0.5\textwidth]{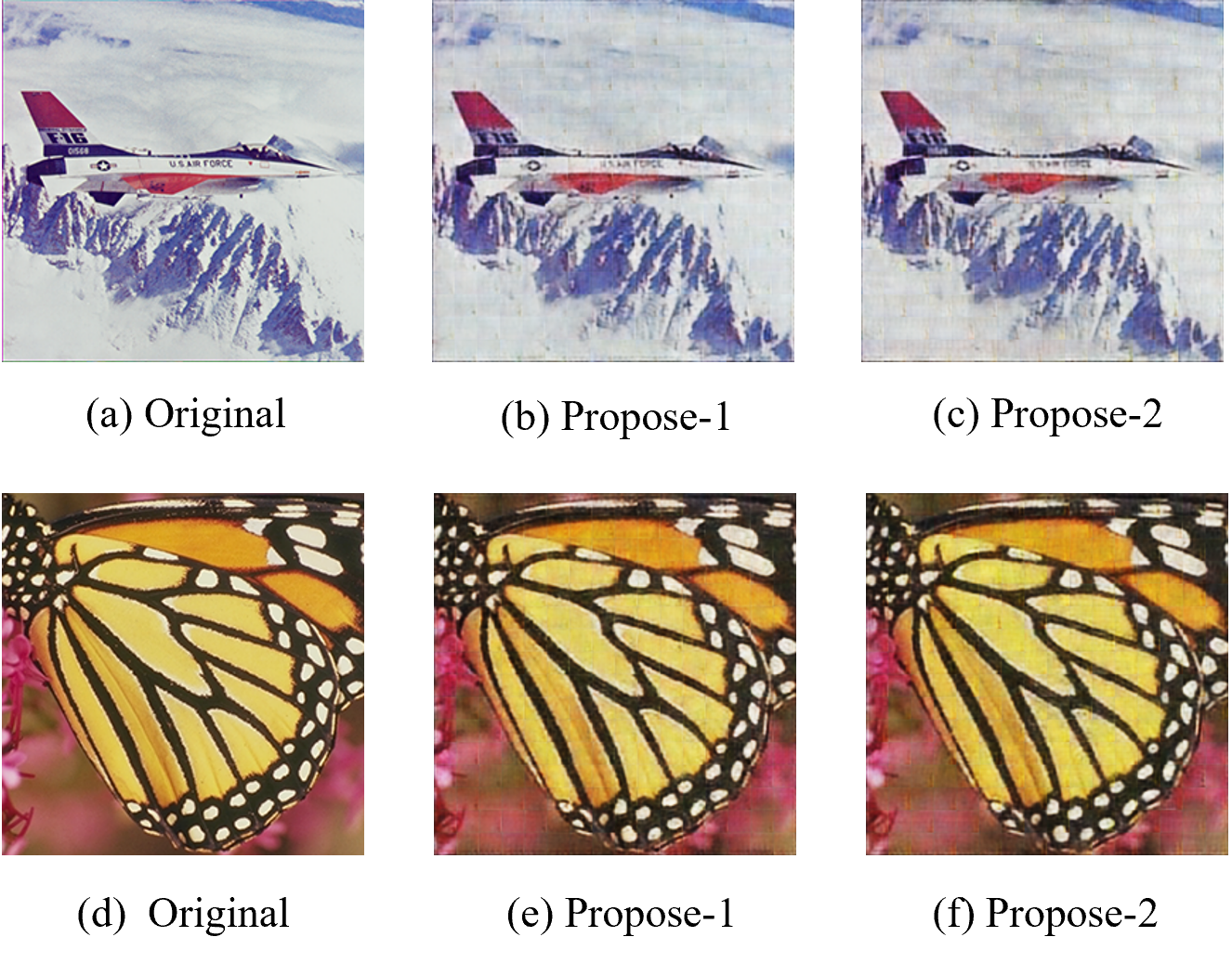}
    \caption{Examples of reconstructions for proposed DJSCC-DNA scheme in $R=1/2$ and $s=128$ for Airplane and Butterfly: For (b)(c)(e)(f), the corresponding PSNR (in $\mathrm{dB}$) and SSIM values are (26.1467,0.8305), (25.2740,0.7910), (25.7561,0.8473) and (24.8674,0.8255)}
\label{fig.Results_big}
\end{figure}

Lastly, we showcase the performance of our schemes on larger image sizes. The original image size is $256\times 256\times 3$. As the model can only train images of size $32\times 32\times 3$, we partition the image into non-overlapping $32\times 32\times 3$ small images, and then append $16$ blank pixels around each small image and subdivide them into smaller images again. All small images are encoded and decoded using the proposed DJSCC-DNA schemes ($\gamma_{\mathrm{train}}=\gamma_{\mathrm{test}}=0.50\%$), and then reassembled after averaging the two stitching results. This process effectively mitigates the noticeable edge-cutting issue caused by block processing. The restored image is shown in Fig.~\ref{fig.Results_big}, with detailed PSNR and SSIM data provided in the caption.

\section{Conclusions}

In this paper, we proposed a novel approach for DNA storage, called DJSCC-DNA, which employs a deep learning technique to effectuate JSCC for DNA image storage. It outperforms conventional neural network-based schemes and effectively administers biological constraints, facilitating a comprehensive system for image storage in DNA.

The innovativeness of the proposed DJSCC-DNA lies in the following areas: We incorporate deep learning into DNA storage and express the encoding and decoding processes through a CNN. We integrate DNA PCR into the network architecture to enhance error resilience. Additionally, we revamp the loss function to strike a balance between image restoration quality and biological prerequisites for model optimization. The simulation outcomes reveal that the suggested scheme efficiently reconstructs high-quality images, surpassing conventional neural network-based schemes in terms of PSNR and SSIM. Moreover, our loss function efficaciously governs homopolymer run-length and GC content of the nucleotide sequence.

As for future endeavors, several potential areas of exploration present themselves: probing the capacity of the DJSCC technique on other forms of DNA storage beyond image data, such as audio, video, and textual data; developing more sophisticated and efficient deep learning algorithms that can augment the performance of DJSCC-DNA and manage larger data sets. With the continued evolution of the DNA storage and information processing domain, a plethora of intriguing research opportunities await discovery.

\bibliographystyle{ieeetr}	
\bibliography{jscc-dna-bib}

\end{document}